\documentclass[aps,pra,showpacs,showkeys,superscriptaddress,groupedaddress,10pt,preprintnumbers,reprint]{revtex4-1}

\usepackage[utf8]{inputenc}	
\usepackage[T1]{fontenc}
\usepackage[english]{babel}

\usepackage{amsmath}
\usepackage{amsfonts}
\usepackage{latexsym}
\usepackage{amssymb} 
\usepackage{MnSymbol} 
\usepackage{bbm} 
\usepackage{textcomp}

\usepackage[pdftex]{graphicx} 
\graphicspath{{./pictures/}}

\usepackage[usenames]{color}

\newcommand{\g}[1]{{\bf #1}}

\newcommand{\ug}[0]{{UGe$_{2}$}}

\begin{document}

\preprint{\scriptsize \it It's an ArXiv copy of the paper published in Journal of Magnetism and Magnetic Materials doi:10.1016/j.jmmm.2015.07.017}

\title{Tricritical wings in UGe$_2$: A microscopic interpretation}
\date{\today}

\author{Marcin Abram}
\email{marcin.abram@uj.edu.pl}
\affiliation{Marian Smoluchowski Institute of Physics, Jagiellonian University, \L{}ojasiewicza 11, PL-30-348 Krak\'ow, Poland}

\author{Marcin M. Wysoki\'nski}
\email{marcin.wysokinski@uj.edu.pl}
\affiliation{Marian Smoluchowski Institute of Physics, Jagiellonian University, \L{}ojasiewicza 11, PL-30-348 Krak\'ow, Poland}

\author{J\'ozef Spa\l ek}
\email{ufspalek@if.uj.edu.pl}
\affiliation{Marian Smoluchowski Institute of Physics, Jagiellonian University, \L{}ojasiewicza 11, PL-30-348 Krak\'ow, Poland}

\date{11 July 2015}

\begin{abstract}
In the present work we analyze the second 
order transition line that connect the tricritical point 
and the quantum critical ending point on the 
temperature--magnetic-field plane in UGe$_2$. For the 
microscopic modeling we employ the Anderson lattice model 
recently shown to provide a fairly complete description of the full magnetic phase 
diagram of UGe$_2$ including all the criticalities. The shape of the so-called tricritical wings, 
i.e. surfaces of the first-order transitions, previously reported by us to 
quantitatively agree with the experimental data, is investigated here with respect 
to the change of the total filling and the Land\'e factor for $f$ electrons which can differ 
from the free electron value. The analysis of the total filling dependence demonstrates 
sensitivity of our prediction when the respective positions of the critical ending point 
at the metamagnetic transition and tricritical point are mismatched as compared to the experiment. 
\end{abstract}

\pacs{71.27.+a, 75.30.Kz, 71.10.-w \hfill doi:10.1016/j.jmmm.2015.07.017}

\maketitle

\section{Motivation and Overview}\label{intro}

\begin{figure} 
  \begin{center}
   \includegraphics[width=0.44\textwidth]{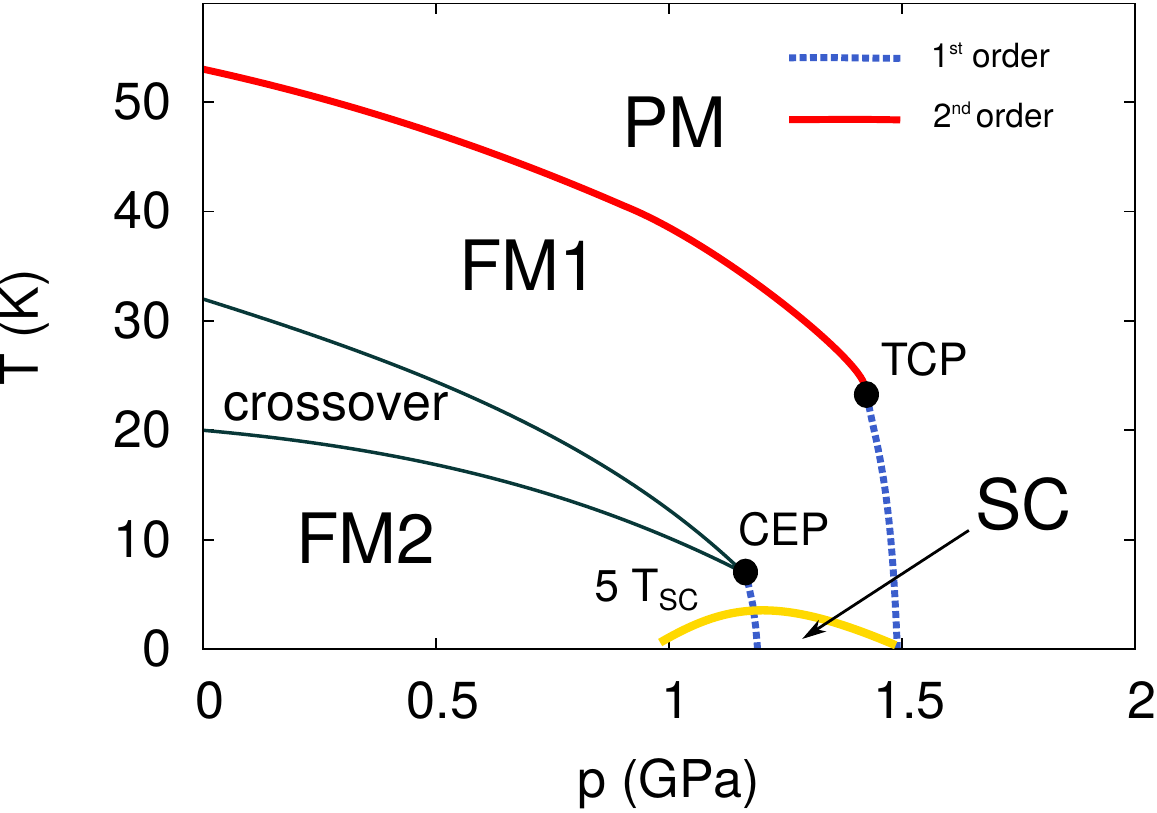}
   \caption{Schematic magnetic phase diagram of \ug\ on pressure--temperature plane drawn on the 
   basis of the experimental results \cite{Taufour2010}.}  
   \label{Fig1}
 \end{center}
 \end{figure}

Quantum critical phenomena have captured general attention  
due to their unique singular properties observed at low 
temperature ($T\rightarrow0$) and near the quantum critical point (QCP)
which is frequently accompanied by the unconventional superconductivity (SC)
\cite{Pfleiderer2009}.
From this perspective, $f$-electron compound \ug\ is a system with phase diagram
comprising coexistence of spin-triplet SC and ferromagnetism (FM)
\cite{Saxena2000,Aoki2012,Aoki2014,Aoki2014a,Huxley2015}, as well as an abundance of 
critical points (CP), either of quantum and classical nature \cite{notes}.
Experimental studies among others have revealed existence of the two characteristic classical CPs 
in the absence of the field (cf. Fig.~\ref{Fig1}):
\emph{(i)} the critical ending point (CEP) at 7K at the metamagnetic transition  
separating strong (FM2) and weak magnetization (FM1) regions \cite{Pfleiderer2002,Taufour2010,Kotegawa2011}, and
\emph{(ii)} the tricritical point (TCP) at the FM to paramagnetic (PM) phase transition located at $T=24$~K.
Additionally, with the applied magnetic field the second order transition line 
starting from TCP can be followed to $T=0$ where it is expected to terminate in a 
quantum critical ending point (QCEP) \cite{Taufour2010,Kotegawa2011}. In effect the magnetic phase boundaries
in \ug\ reflects so-called tricritical wing shape.

Such complex magnetic phase diagram with all the above criticalities, both classical and quantum,
is particularly challenging in terms of theoretical modeling.
One of the first approaches, based on the single-band model describing tricritical wings, 
was the work by Belitz, et al. \cite{Kirkpatrick2005}.
However, the microscopic description
of the magnetic phase diagram with all the CPs including also CEP at the metamagnetic transition, 
as observed in \ug, has been missing
until our recent works  \cite{Rapid1,Rapid2}.

Our analysis is based on the (two-orbital) Anderson
lattice model (ALM) \cite{Rapid1,Rapid2}, often referred to as the periodic Anderson model.
Findings for \ug,\ both from first principle calculations and experiments are:  
the quasi-two-dimensional character of the Fermi surface \cite{Shick2001}, a uniaxial
anisotropy for magnetization \cite{Huxley2001}, U--U interatomic distance above the so-called Hill limit \cite{Pfleiderer2009},
and the paramagnetic moment per U atom different from that expected for either $f^3$ or $f^2$ atomic configurations \cite{Saxena2000,Kernavanois2001}.
We show in the following that all of these findings can be coherently explained within
our two-orbital model starting with originally localized $f$-states and subsequently being 
strongly hybridized with the conduction ($c$)
band states on a two dimensional lattice and with the applied magnetic field
accounted for by the Zeeman term only.

Ferromagnetic order in our model arises from effect of competing hybridization 
and the $f$--$f$ interatomic Coulomb repulsion. The emergence of two distinct ferromagnetic phases 
is in our model driven by the changing topology of the Fermi surface \cite{Doradzinski1997,Doradzinski1998,
Howczak2013,Kubo2013} which in turn is induced by a relative motion of hybridized and 
spin split subbands with the increasing $f$-$c$ hybridization. 
The results obtained from such picture \cite{Rapid1} qualitatively agree with the majority of \ug\ 
magnetic and electronic properties, as seen in neutron scattering \cite{Kernavanois2001}, 
de Haas-van Alphen oscillations \cite{Terashima2001, Settai2002}, and magnetization
measurements \cite{Pfleiderer2002}. Also, a semi-metallic character of the weak FM1 phase is
supported by the band-structure calculations \cite{Samsel2011}. 
A similar idea concerning the emergence of two distinct FM phases in \ug\ was also 
obtained earlier within the phenomenological picture based on the 
Stoner theory incorporating a two-peak structure of the density of states in a single band \cite{Sandeman2003}.
In brief,
our microscopic model extended to the case of $T>0$ \cite{Rapid2}  
describes well emergence of all CPs on the magnetic phase diagram of \ug\ 
\cite{Pfleiderer2002, Taufour2010,Kotegawa2011} in the semiquantitative manner \cite{Rapid2}.
Here we compare in detail our results with the experimental 
data, namely predicted second order transition line joining the TCP with the corresponding QCEP.
In particular, we determine the influence of the following factors:
\emph{(i)} the total band filling $n$, 
and \emph{(ii)} the value of the Land\'e factor $g_f$ for $f$ states, on the position of this second-order line. 
The influence of factor (i) has the following importance.
For exemplary filling  $n=1.6$ we have shown \cite{Rapid2} that the relative 
position of the TCP and CEP (cf. Fig.\ref{Fig1}) is the same as that seen in the experiments \cite{Taufour2010,Kotegawa2011}.
The important question is whether such a mutual alignment of those two critical points is necessary to achieve 
a good fit and to what extent the proper curvature of the line joining TCP and QCEP is robust with respect to the selected band filling.
The discussion of the dependence on (ii) has its justification in the not fully resolved
nature of magnetism in heavy-fermion systems in general and \ug\ in particular.
Although it is assumed and widely accepted to be fully itinerant \cite{Saxena2000},
there is evidence for a partially localized contribution \cite{Samsel2011,Troc2012}.
In such case, the influence of the orbital effects and their coupling to the spin
should have an influence on $g_f$ value.

\section{Model and approach}
We begin with the orbitally nondegenerate Anderson-latice model (ALM)
on square lattice and with applied magnetic field 
accounted for via the Zeeman splitting
(i.e., with the effective field is $h \equiv \frac{1}{2} g \mu_0\mu_B H$),
so that the starting Hamiltonian is
\begin{equation}
 \begin{split}
 \mathcal{\hat H}_0&={\sum_{\g i,\,\g j,\,\sigma}}'  t_{\g i \g j}\hat c_{\g i\sigma}^\dagger\hat c_{\g j\sigma}
-\sum_{\g i,\sigma}\sigma h\ \hat n^c_{\g i\sigma}\\
&+ \sum_{\g i,\,\sigma}(\epsilon_f-\frac{g_f}{g}\sigma h)\hat n^f_{\g i\sigma}+ 
U\sum_{\g i} \hat n^f_{\g i\uparrow} \hat n^f_{\g i\downarrow}\\
&+ V\sum_{\g i,\,\sigma}(\hat f_{\g i\sigma}^\dagger
\hat c_{\g i\sigma}+\hat c_{\g i\sigma}^\dagger\hat f_{\g i\sigma}),\label{H1}
\end{split}
\end{equation}
where the onsite hybridization is of magnitude ${V<0}$ and the Land\'e factor for 
$f$ electrons is $g_f$ (the free electron value is $g=2$).
The model describes a two-orbital system with the conduction $(c)$ band arising from the nearest ($t$) and
the second-nearest ($t'$) neighbor hoppings,
and the strong $f$--$f$ Coulomb interaction is of magnitude $U$.
If it is not stated otherwise, we set $t'=0.25|t|$, $U=5|t|$, $\epsilon_f=-3|t|$, $g_f=g=2$, and 
${n\equiv\sum_{\sigma} \langle\hat n^c_{\g i\sigma}+\hat n^f_{\g i\sigma}\rangle=1.6}$.  

We also add to the Hamiltonian (\ref{H1}) the usual term with the chemical potential $\mu$, i.e.,
\begin{equation}
 \mathcal{\hat H}\equiv \mathcal{\hat H}_0-\mu\sum_{\g i,\sigma} (\hat n^c_{\g i\sigma}+\hat n^f_{\g i\sigma}).
\end{equation}
The model is solved here by means of \emph{statistically consistent Gutzwiller approximation} (SGA)
\cite{Rice1985,Fazekas1987,SGA}. 
The method was successfully applied to the number of problems \cite{Abram2013,Wysokinski2014}.
It is characterized with the physical transparency and flexibility, that it could be also incorporated into other
methods such as EDABI \cite{Kadzielawa2013, Kadzielawa2014}.

We introduce the Gutzwiller projection acting onto uncorrelated wave function $|\psi_0\rangle$ 
in the following manner 
\begin{equation}
 |\psi_G\rangle=\prod_{\g i} P_{G;\,{\bf i}}\,|\psi_0\rangle, \label{H}
\end{equation}
where $|\psi_G\rangle$ is the wave function of the correlated ground state.
In effect, we map many-particle correlated Hamiltonian (\ref{H1}) onto an
effective single-particle Hamiltonian $\mathcal{\hat H}_{SGA}$ acting on uncorrelated wave function 
$|\psi_0\rangle$, that, after taking the space Fourier transform, is as follows
\begin{multline}
 \mathcal{\hat H}_{SGA} \equiv \hat\Psi_{\g{k}\sigma}^\dagger 
 \begin{pmatrix}
 \epsilon_{\g{k}}^{c}-\sigma h-\mu&\sqrt{q_\sigma}V \vspace{3pt}\\
 \sqrt{q_\sigma}V& \epsilon_{f}-\sigma \frac{g_f}{g} h-\mu   \\ 
\end{pmatrix}
\hat\Psi_{\g{k}\sigma} \\
 -\lambda^f_n \Big( \sum_{\g{k},\,\sigma}\hat n^f_{\g{k}\sigma}- \Lambda n_f \Big)
 -\lambda^f_m \Big( \sum_{\g{k},\,\sigma}\sigma\hat n^f_{\g{k}\sigma} -  \Lambda m_f \Big) \vspace{3pt}\\
+\Lambda Ud^2 
\label{HSGA}
\end{multline}
where $\hat\Psi_{\g{k}\sigma}^\dagger\equiv(\hat c_{\g{k},\sigma}^\dagger, \hat f_{\g{k},\sigma}^\dagger)$,
$\Lambda$ is the number of the system sites, $q_\sigma$ is the hybridization narrowing factor,
and $d^2 \equiv \langle \hat n_{{\bf i}\uparrow}^f \hat n_{{\bf i}\downarrow}^f \rangle_0$ \cite{Wysokinski2014}.
Necessary constraints for the $f$ electron number and their magnetic moment \cite{SGA} are incorporated
by means of the Lagrange multipliers $\lambda^f_n$ and $\lambda^f_m$, respectively.
Hamiltonian (\ref{HSGA}) can be straightforwardly diagonalized with the resulting four eigenvalues $\{E_{\g k\sigma}^b\}$ labeled with 
the spin ($\sigma = \pm 1$) and hybridized-band ($b=\pm 1$) indices.
For $T>0$ we construct a generalized Landau grand-potential functional according to
\begin{equation}
\begin{split}
\frac{ \mathcal{F}}{\Lambda}  = & -\frac{1}{\Lambda\beta} \sum_{{\bf k} \sigma b} \ln[1+e^{-
\beta E_{{\bf k}\sigma}^b}]\\
&+ (\lambda_{n}^fn_f+\lambda_{m}^fm_f+Ud^2),
\label{5}
\end{split}
\end{equation}
which is next minimized with respect to the set of following parameters assembled to
a vector, $\vec \lambda\equiv\{d,n_f,m_f,\lambda_n^f,\lambda_m^f\}$.
Additionally, we adjust self-consistently the chemical potential
from the condition of fixing the
total number of electrons,
${n=1/\Lambda\sum_{{\bf k}b\sigma}  f ( E_{{\bf k} \sigma}^{b} )}$,
where $f(E)$ is the Fermi-Dirac function. 
Finally the ground state energy is defined by,
\begin{equation}
 E_G=\mathcal{F} \big|_0 + \mu N,
\end{equation}
where $\mathcal{F}\big|_0$ means that the value of $\mathcal{F}$ is taken at minimum for the parameters $\vec \lambda$.

\section{Results and discussion}

We start our analysis with the discussion on the proper assignment of the physical units to 
the microscopic parameters provided so far in dimensionless
units (i.e. scaled by $|t|$) to make the quantitative comparison with the observed \ug\ characteristics.
To do so, we have adjusted them \cite{Rapid2} by matching the relative 
positions of the two classical CPs: TCP and CEP at the ferromagnetic transition, as well as attributing the experimentally measured critical temperatures. 
Matching the results in physical units by fixing the position of two critical points we would call
a {\it strong fitting}, whereas by fixing the position of just a single of them a {\it weak fitting}.

In our previous works \cite{Rapid1,Rapid2} we have found that for the total filling $n=1.6$ we could coherently and quantitatively 
describe the \ug\ phase diagram. Although, there is no direct experimental evidence in \ug\ for choosing this particular 
filling, we have first matched for chosen $n$ TCP and CEP temperatures according to the experiment \cite{Rapid2} -- strong fitting condition, and second, we have
verified our prediction obtaining agreement of the second-order transition
line joining TCP with QCEP \cite{Kotegawa2011} with that measured.
Additionally, the comparison has provided among others the estimate
of the QCEP appearance about  $30$~T, i.e., higher than that suggested in Ref.~\cite{Kotegawa2011} which is $18$~T.

\begin{figure} 
   \begin{center}
   \includegraphics[width=0.5\textwidth]{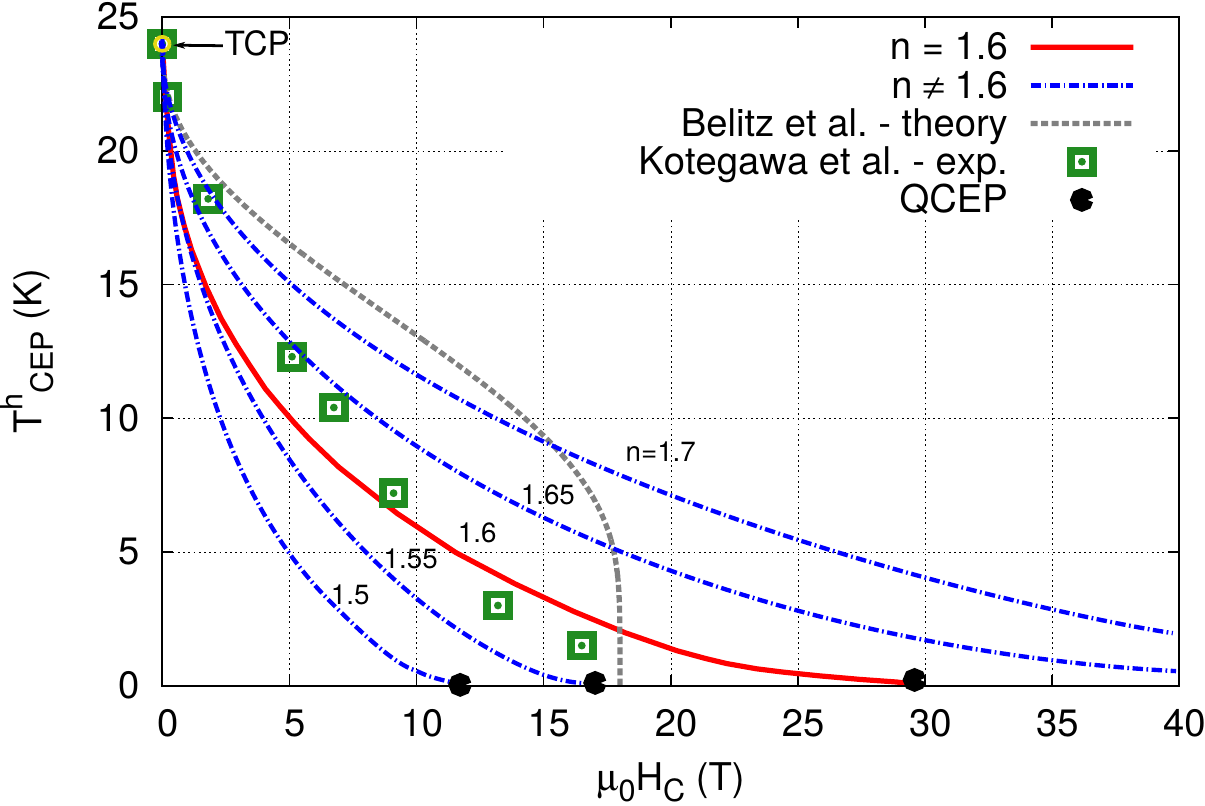}
   \caption{The second-order transition line joining TCP and QCEP for selected values
   of total filling, $n$. Solid, red curve is replotted from the results reported in Ref. \cite{Rapid2}.
   Experimental points are extracted from pictures in Ref. \cite{Kotegawa2011}. Grey, dashed curve predicted by the 
   theory formulated by Belitz, et al. \cite{Kirkpatrick2005} is shown for comparison (see main text). }
   \label{Fig2}
 \end{center}
 \end{figure}

A natural question arises if this test is sensitive to the choice of $n$. We show in Fig.~\ref{Fig2} 
that the second-order transition line joining TCP and QCEP determined for a slightly different total filling than $n=1.6$
deviates significantly from the trend of the experimental data \cite{Kotegawa2011}. Hence 
indeed, the comparison is very sensitive to the choice of $n$.
Thus, together with equally sensitive adjustment of $T_{TCP}$ and $T_{CEP}$ with respect to the choice of $n$ \cite{Rapid2}, 
it is unlikely that our excellent agreement for the single value of parameter $n$ is fortuitous.

For the sake of completeness and reference to other related works we include 
in Fig.~\ref{Fig2}  the (dashed) curve predicted by the one 
of the most successful approaches describing the general 
tricritical behavior in itinerant magnetic systems \cite{Kirkpatrick2005,Kirkpatrick2012}. 
In that procedure, the necessary inputs are the positions of the two CPs, namely TCP and QCEP, 
leading in fact to the  {\it strong fitting}, but with different pair of CPs.
However, such fitting can be associated with an error as the position of the QCEP, 
in contrast to TCP and CEP, is not experimentally determined but
only extrapolated to 18~T, following Ref.~\cite{Kotegawa2011}. Note that
this condition in our modeling is satisfied for the total filling around $n\simeq1.55$ (cf. Fig.~\ref{Fig2}) 
if we take the extrapolated value of the critical field. In this case, the comparison with 
results for \ug\ \cite{Kotegawa2011} is worse
then e.g. for $n=1.6$ and the temperature of the CEP is much lower than that determined in experiments 
\cite{Pfleiderer2002}. 

It is worth mentioning that the model employed by us belongs to the  
class discussed earlier by Kirkpatrick and Belitz \cite{Kirkpatrick2012} to reflect
the generic tricriticality in the case of metallic magnets.
Namely, systems in which the conduction electrons are not 
a source of the magnetism themselves, but are couple to the magnetically ordered localized electrons in a second band.
The origin of the first-order transition at low temperature described within the mean-field theory developed in the Refs. 
\cite{Kirkpatrick2005,Kirkpatrick2012} is based on the effect of  
the soft fermionic modes coupled to the magnetization fluctuations, and thus differs
from our approach. Here the mechanism for ferromagnetism is due to the coupling
of the conduction electrons with localized $f$ states by hybridizing with them and competing
with the ${f{\rm{-}}f}$ Coulomb interaction. This competition in the Stoner-like manner 
induces phase transitions associated with the abrupt changes of the Fermi surface topology.

The simplest verification of our analysis can be carried out
by means of chemical alloying, i.e., by changing the electron concentration in the system.
However, the lack of known isostructural compounds to \ug\ may be an apparent obstacle for such test.
Though, the determination of the tail of the 2nd order line joining TCP and QCEP for the field larger than 16 T should provide
an insight on the issue whether our model correctly predicts the appearance of QCEP around 30 T \cite{Rapid2}.  

    \begin{figure} 
  \begin{center}
   \includegraphics[width=0.45\textwidth]{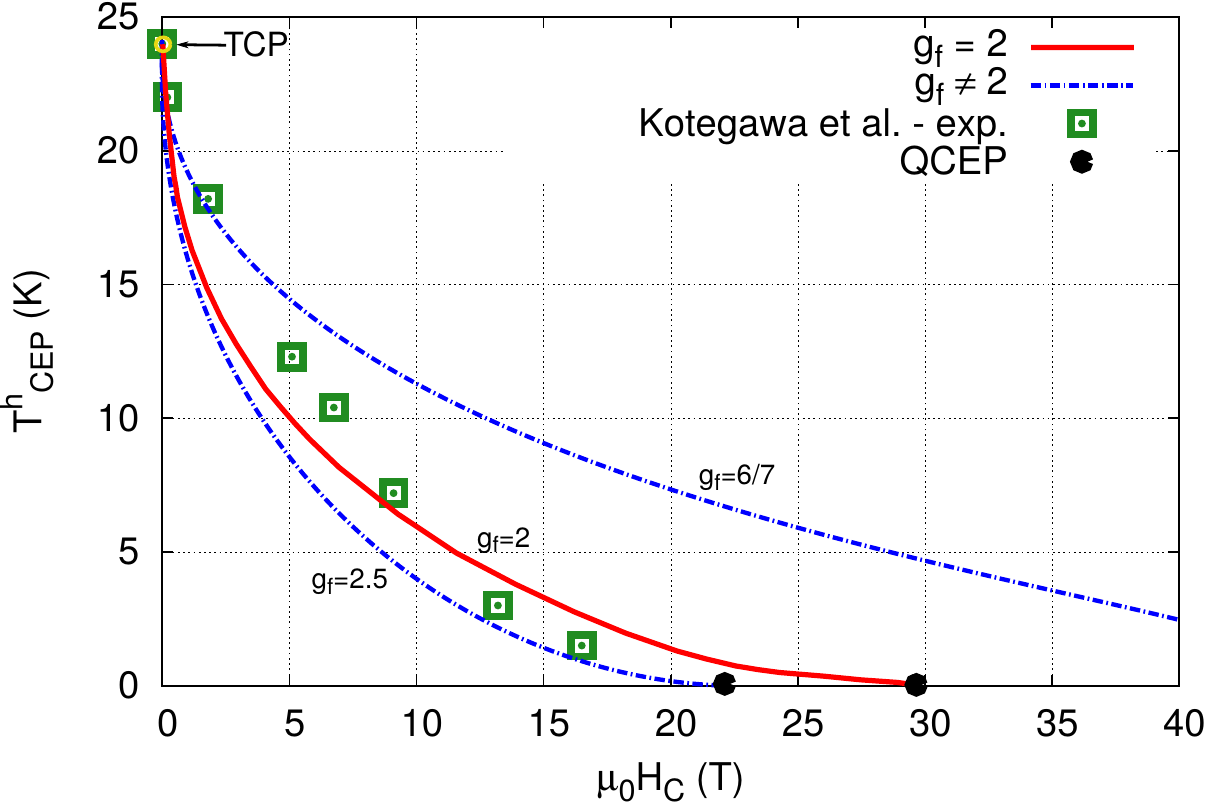}
   \caption{The second-order transition line joining TCP and QCEP for selected values
   of Land\'e factor for $f$ electrons, $g_f$. Solid curve for $g_f = 2$ is replotted from the results reported 
   in Ref. \cite{Rapid2}. Experimental points are extracted from Ref. \cite{Kotegawa2011}.}
    \label{Fig3}
\end{center}
 \end{figure}

If our model is to be used to understand the magnetism of 
other ferromagnetic superconductors: URhGe \cite{Aoki2001} and UCoGe 
\cite{Huy2007}, it would provide a perfect testing ground of our model 
as those compounds have been frequently studied by means of the chemical substitution 
\cite{Pfleiderer2009,Aoki2012,Aoki2014,Aoki2014a,Huxley2015}.

Finally, we provide a brief analysis of the impact of the Land\'e factor value for $f$ electrons, $g_f$,
i.e., in the situation when the $z$ component of the total spin of the system does not commute with
Hamiltonian. In Fig.~\ref{Fig3} we present the curves for three different values of $g_f$.
The curve for $g_f=2$, is plotted as the reference curve and is based on the results of Ref. \cite{Rapid2}.
Value of $g_f$ is not known for \ug\ and generally, 
for complex compounds is has a tensor character which depends on the magnitude of the spin-orbit coupling. 
For that reason we restrict our discussion to the comparison when $g_f$ 
is equal to the free electron value $g_f=2$, and subsequently when is lower and higher 
(cf. Fig.\ref{Fig3}) \cite{Quinet2004207}.
Specifically, the lower value of Land\'e factor $g_f=6/7$ is
motivated by that for the Ce-based compounds, where it can be derived for  the spin $S=1/2$ and 
angular momentum $L=3$, oriented antiparallel and where, strictly speaking, our model is also generally valid, as
long as we do not account for the orbital degeneracy of $f$ states of the uranium-based materials.
As presented in Fig. \ref{Fig3}, it seems that any value of  
$g$ factor for $f$ states which deviates considerably from the free-electron value, 
provides much worse agreement with experimental data \cite{Kotegawa2011}.
In conclusion, due to predominantly itinerant nature of $f$ electrons in \ug\ \cite{Troc2012}, it is very
likely that any crystal-field derived multiplet structure is washed out and hence the value $g_f\simeq2$ should be
regarded as realistic value. Nevertheless, the problem of double localized--itinerant nature
of $f$-electrons \cite{Samsel2011,Troc2012} may arise 
as the system evolves with the increasing temperature, in comparison to the pressure evolution
at low temperature studied in detail here.

   \begin{figure}[b]
  \begin{center}
   \includegraphics[width=0.5\textwidth]{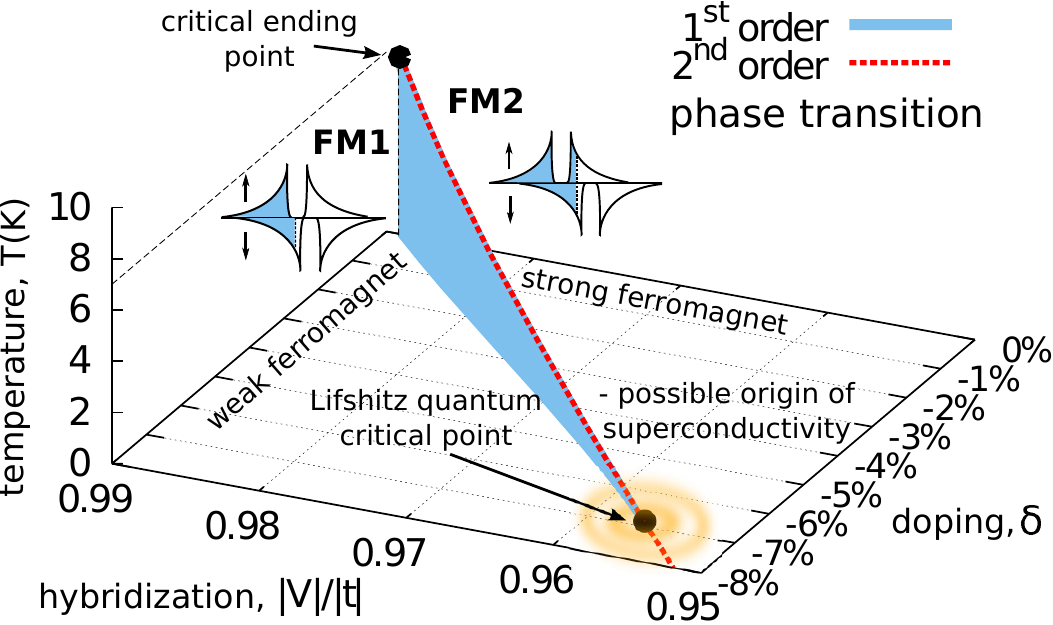}
   \caption{Evolution of the critical temperature of CEP towards Lifshitz QCP driven by 
  changing total filling, $n$ and hybridization, $V$ reproduced after \cite{Rapid2}.}
   \label{Fig4}
 \end{center}
 \end{figure}
 
\section{Outlook}

In the present work we have employed the Anderson lattice model \cite{Rapid1,Rapid2}
to provide a fairly complete description of the magnetic phase diagram (p--T--h profiles) of \ug\
including all the criticalities for this compound.
In particular, we study the effect of the choice of the total filling on
the quality of the fit, based on our model, to the experimental data \cite{Kotegawa2011} 
concerning the second-order transition line joining the critical points TCP and QCEP.
We have found that our prediction is very sensitive to the change of $n$, which leads also
to a mismatch of critical temperatures of TCP and CEP at at the metamagnetic transition as compared to the experiment.
We infer from this result that our excellent agreement for the single value of $n$ is unlikely fortuitous.
We have also analyzed the effect of the Land\'e factor $g_f$ value for $f$ electrons.
In this case, any sizable deviation from the free electron value $g_f=2$ shifts the theoretical curves away from the experimental points. 
Thus treating $f$ electrons as truly itinerant electrons in \ug\ seems to be fully justified.

Our final remark addresses the problem of the spin-triplet superconductivity (SC) origin occurring in \ug\ \cite{Pfleiderer2009, Saxena2000}. 
We have predicted in our previous work \cite{Rapid2} the appearance of QCP in the vicinity 
of the SC dome.
It have been proposed that that CEP (cf. Fig.~\ref{Fig1}) at the metamagnetic
phase boundary can be followed down to the $T=0$ by
changing both the electron concentration and the hybridization magnitude $|V|$ (cf. Fig.~\ref{Fig4}).
The proposed quantum critical point is of Lifshitz type as it 
separates states with two distinct Fermi-surface topologies. Quantum critical fluctuations
or the residual $f$-$f$ Hund's rule interaction (neglected here) can become the possible source of the spin-triplet
superconductivity \cite{Zegrodnik2012, Zegrodnik2013, Zegrodnik2013JPCM, Zegrodnik2014, spa}.
A detailed and quantitative discussion of the pairing requires a separate analysis.

{\it Acknowledgements.} The work was supported by 
the National Science Centre (NCN) under the Grant MAESTRO, No. DEC-2012/04/A/ST3/00342.

 \bibliographystyle{prsty}

\begin{thebibliography}{10}
\bibitem{notes}
\rm{Coexistence} of FM and SC was observed for the first time by
  A.~Kolodziejczyk for the weak itinerant ferromagnets Y$_9$Co$_7$
  \cite{kolodziejczyk1,kolodziejczyk2,kolodziejczyk3}. However, SC was found
  there to be of spin-singlet nature and strongly competing with FM, in
  contrast to the UGe$_2$ case, where it is of the spin-triplet character and
  its coexsitence with FM seems to be strongly cooperative as e.g. both phases
  disappear at the same pressure.
\end{thebibliography}

\end{document}